\documentstyle[aps,12pt]{revtex}
\newcommand{\p}{\partial}

\renewcommand{\d }{{\rm d}}
\newcommand{\e}{{\rm e}}

\begin{document}
\title{The Dynamics of Thermodynamics \footnote{Presented at the Workshop
on Time-Reversal Symmetry in Dynamical systems, Warwick, U.K., 9th--20th
December 1996.}}
\author{J. Kumi\v{c}\'{a}k}
\address{Department of Thermodynamics, Technical University, \\
041 87 Ko\v{s}ice, Slovakia.}
\author{X. de Hemptinne}
\address{Department of Chemistry, Catholic University of Leuven\\
B-3001 Heverlee, Belgium}
\maketitle
\vspace{10mm}

\begin{abstract}
Thermodynamic relations are derived from first principles of mechanics for
non-equilibrium processes. Since the key role herein is played by the law
of increase of entropy, the latter is analyzed at first. It is shown that
its derivation for isolated systems does not allow one to say too much
about thermodynamic properties in non-equilibrium conditions. From there
on, the notion of quasi-isolated systems is introduced, for which it is
possible to describe the state of macroscopic systems by a collection of
intensive variables. The latter are the differentials of the entropy with
respect to the given set of extensive constraints. Arguments are developed
showing that the dynamics of irreversible relaxation of non-equilibrium
systems consists of two qualitatively different steps. The theory is
applied and verified by describing transport processes and calculating the
corresponding coefficients. Structure formation and transition to
turbulence are equally formalized in the thermodynamic framework.

\end{abstract}

\section{Introduction}
Contemporary mathematical literature assigns an abstract meaning to the
concept of {\em dynamical systems\/}. They are most often considered as
substrates for particular problems in applied mathematics. Physics, by
contrast, prefers thinking of dynamical systems as being macroscopic
objects with time-de\-pen\-dent properties, consisting of microscopic
particles responding to the laws of mechanics. The goal here is to
establish links between observable macroscopic properties and the more or
less predictable motion or trajectories of the system's elements. By
quoting formally {\em fluid dynamics\/} and {\em statistical physics\/} in
the workshop's announcement as topics to be covered, the physicist's
definition is implicitly included in the general debate. This implies
possible confrontation with the experimental reality for validation of the
conclusions. The present review has the latter aspect of the problem of
dynamical systems in mind.

The general discussion concerns time-reversal properties. With physical
systems, not considering (or excluding) possible external action, the
microscopic laws of mechanics governing the motion of the particles are
Hamiltonian and therefore time-reversible. Macroscopic time-dependent
properties show however a high degree of irreversibility. Apparent
contradiction between the symmetry properties of the microscopic laws and
the experimentally verified macroscopic behaviour of dynamical systems
remains a controversial subject of debate.

Macroscopic time-dependent phenomena shown by physical systems include
relaxation of disturbances, transport effects like friction and others and
bifurcations to more or less ordered dissipative structures. The last item
has been quoted explicitly in the list of dynamical phenomena to be
covered by the present general discussion. Considering the many
connections prevailing between this item and the other time-dependent
effects mentioned above, a general discussion including relaxation and
transport effects is justified.

In this review it will be proven that next to the internal Hamiltonian
laws of motion, observable physical systems rely on interactions with the
external world for their global dynamics. Depending on the conditions, the
step that controls the relevant time-dependent phenomenon and defines the
observed degree of time symmetry may depend either on internal or external
effects. In all cases a correct formalism is required for expressing the
properties of the surroundings and their relation to that of the system
itself. This is achieved by extending the principles of thermodynamics to
conditions removed from equilibrium. A review of paradigms and assumptions
supporting the time-reversal symmetry problem in statistical physics will
be given first. They will be critically analyzed and a new approach will
be suggested based on analysis of Joule's experiment. The necessary
thermodynamic tools will then be developed. The thermodynamic formalism
will be applied for predicting and discussing selected topics in fluid
dynamics. In a first step, the transport coefficients are considered.
Structure formation and transition to turbulence are developed from there
on.

\section{Controversies}
Distortion of macroscopic systems initiates spontaneous and irreversible
pro\-cesses tending to restore the previous state of equilibrium or
possibly to establish a new one. This fact is most often taken to be a
genuine property of conservative Hamiltonian systems. Conflict between the
time asymmetric behaviour of macroscopic relaxation and the strict
reversible nature of microscopic dynamics of conservative systems has been
the subject of discussions over nearly a century \cite{leb:73,lebA:93}.

Results published in recent decennia in applied mathematics concerning
time dependent transformation of systems where the number of identical
elements tends to infinity at constant density \cite{sinai:76,cornfeld:82}
have been a stimulus for trying to solve the irreversibility paradox. The
arguments are related to the mathematical property of mixing. This is
meant to express dissemination of the system's parameters throughout the
available space towards homogeneous and statistically independent
distributions. It associates irreversibility to an infinite Poincar\'{e}
recurrence time for most initial fluctuations. Furthermore, progress
obtained in characterizing deterministic chaotic motion (Lyapounov
exponents) spurs theoretical research towards relating the relevant
numbers to transport properties associated with irreversible relaxation
dynamics. In this context chaotic scattering of particles on hard disks
(Lorentz gas) is used as a model for describing irreversible diffusion in
agreement with reversible microscopic dynamics
\cite{spohn:80,gasp:92,gasp:95,dorf:95}.

Using the language of functional analysis one arrives at a rigourous
formulation of the fundamental difference between reversible and
irreversible evolutions. The reversible evolution of dynamical systems is
expressed by a group of isometric operators, whereas the irreversible one
is described by a semigroup of strictly contractive ones. As
microscopic laws are time-reversible, the evolution they cause must be
formulated by isometric groups. By contrast, macroscopic irreversible
processes can be described only by contractive semigroups. We are thus
faced with the question of how this transition from a group to a semigroup
can be justified. The presently most popular answer is proposed by the
Brussels school associated with the name of I.~Prigogine~\cite{MPC}. This
school's interpretation rejects formally the role of loss of information
in deriving the second law of thermodynamics. It has been subjected to
criticism in~\cite{Kum87} on the basis of mathematical results of
G.~Braunss~\cite{Braunss}.

Aforementioned research trends assume explicitly that irreversible
processes would occur in systems isolated from the environment. The
statistical properties of the time dependent random forces acting on the
particles are supposed to be {\em completely\/} determined by the initial
conditions and by the dynamics of the system \cite{spohn:80}. Rescaling is
said to make phenomenological parameters (e.g. viscous drag controlling
the motion of Brownian particles) converging to genuine irreversible
properties of the system.  Dissipative coupling with a reservoir is
therefore explicitly rejected, as ``artificial and unnecessary''
\cite{lebA:93,gasp:92}.

Contrasting with the latter, some authors still insist on the unavoidable
interaction of macroscopic systems with their environment acting as a
reservoir or heat bath \cite{posch:88,evans:90}. Stress is laid on the
environment where the fluctuating forces to be introduced into the
equations of motion of the system of interest come from. Dissipation
arises from back-reaction of the environment to the evolution of the
system \cite{linden:90,xh:92}. As an example, concerning viscous drag as
mentioned above, the relevant dissipative force is said to originate from
interaction of the Brownian particle with the surrounding fluid acting as
a reservoir. The time averaged value of this force gives rise to Stokes'
law, on which fluctuations are superimposed.

To anticipate and prevent future ambiguities, we specify the meaning to be
given to some important key-words frequently used in the discussion to
follow.

By {\em isolated\/} systems are meant systems the dynamics of which is
defined in a unique way by time-independent Hamiltonians. They are
obviously conservative. Since we are focussing next on systems that are
not perfectly isolated, we need to give them an appropriate name. A system
will be called {\em quasi-isolated\/} if it exchanges energy but not
matter with its surroundings.

{\em Equilibrium\/} refers to any macrostate that is {\em stationary\/}
(i.e.\ time-independent) under the given macroscopic constraints. As an
example we may consider that this property holds for a gas contained
within walls at different temperatures when it has reached stationary
conditions. By {\em approach to equilibrium\/} we denote any observable
macroscopic evolution towards equilibrium.

\section{Isolated dynamical systems}
\subsection{The Problem}
\label{Problem}
Consider a system of $N$ identical structureless classical particles
constrained to move in volume $V$. The state of the $i$-th particle is in
unique way given by the vector of generalized coordinate $\vec q_i$ and by
the vector of generalized momentum $\vec p_i$. Expression $(q,p)$ will
denote the state of the collection of $N$ particles, or a point in the
system's phase space $\Gamma$. This system's dynamics is described by
Hamilton's equations with the time-independent Hamiltonian
$H(q,p)$~\cite{Mechanika}. The action of the walls is represented by a
high repulsive potential (a possible choice being an infinite
potential barrier). The solution $(q(t),p(t))$ to these equations for some
initial state $(q(0),p(0))$ can be written using a group of dynamical
evolution operators $T_t$:

\begin{equation}
(q(t),p(t)) = T_t (q(0),p(0)).
\label{T_t}
\end{equation}
It represents the motion of a point along a trajectory in phase space
$\Gamma$.

Investigation of the evolution of an individual system is apparently not
possible and therefore one resorts to statistical methods~\cite{Zubarev}.
One considers an ensemble of equivalent systems. The ensemble properties
are then described by means of a {\em distribution function\/} $f(q,p;t)$,
giving the probability density of finding a system at point $(q,p)$ in
phase space at time $t$~\cite{Landau_Stat}.

From the deterministic description defined by the group of operators
$T_t$, one goes over to the statistical description of evolution of the
distribution function by introducing a group of operators $U_t$ defined
through relation
\begin{equation}
U_t f(q,p) = f(T_{-t}(q,p))\ .
\label{Koopman}
\end{equation}
It is easy to show that the statistic evolution operator $U_t$ is unitary
--- and thus isometric --- if it acts in a Hilbert space ${\cal
L}^2(\Gamma)$ of square-integrable functions defined on phase space
$\Gamma$~\cite{Arnold}. By Stone's theorem there exists then a
self-adjoint operator $L$ ({\em Liouville operator\/}), such that $U_t =
e^{-iLt}$. Using the Poisson bracket notation it can be easily proved that
$L$ can be expressed by means of the Hamiltonian function as $ L f =
i\{H,f\}$.

The {\em Liouville equation\/},
\begin{equation}
i\frac{\p f}{\p t}=L f,
\label{Liouv}
\end{equation}
has for solution: $f(q,p;t) = U_t f(q,p;0) = e^{-iLt} f(q,p;0)$, where
$f(q,p;0) \equiv f(q(0),p(0))$.

Let $\varphi$ denote one of the possible equilibrium functions. Since
$\varphi$ does not change with time, $\p \varphi/\p t = 0$, whence $L
\varphi = 0$. From~(\ref{Liouv}) we have obviously $U_t\varphi=\varphi$.
Now, unitarity of $U_t$ (i.~e.\ the equality $U_t^{\ast}=U_{-t}$) implies,
first of all, invariance of the scalar product in the course of evolution
\begin{equation}
\langle U_tf\,|\,U_tg\rangle = \langle f\,|\,g\rangle\quad \mbox{for any
} f, g \in L^2(\Gamma)\ .
\end{equation}
This reduces to $\|U_tg\|=\|g\|$ = const for $f=g$. Taking now $g=f-
\varphi$, where $f$ is any nonequilibrium distribution function, we obtain
\begin{equation}
\| U_t f - \varphi \| = \mbox{const}\ .
\end{equation}

Hence, approach of mechanical systems to equilibrium cannot be described
statistically as a strong convergence of the distribution function to an
equilibrium function.

\subsection{BBGKY Hierarchy}
\label{BBGKY}
The description of a system on the basis of a distribution function
$f(q,p;t)$ is complete from a statistical point of view. However, solving
equation~(\ref{Liouv}) explicitly is apparently not possible. Since the
distribution function cannot converge strongly to any equilibrium
function, the entropy as defined by Gibbs by relation
\begin{equation}
S_G = - \int_{\Gamma} f(q,p;t)\ \ln f(q,p;t)\ dq\ dp
\label{Gibbs_entropy}
\end{equation}
is equally time-independent~\cite{Zubarev}. This drawback is eliminated by
introducing so-called {\em reduced distribution functions\/} which are
obtained by integrating function $f(q,p;t)$ over a subset of particles.
For example, we obtain an $r$-particle distribution function $f_r(q,p;t)$
by integrating $f(q,p;t)$ over positions and momenta of $N - r$ particles
($0 < r \leq N - 1$). Such a function describes statistical properties of
a subsystem comprised of $r$ particles.

It turns out that subsequent integration as above of Liouville's
equation~(\ref{Liouv}) leads to a system of $N - 1$ integro-differential
equations for a set of $N - 1$ reduced distribution
functions. The equations have the property that the equation for $f_r$ ($r
< N$) contains always function $f_{r+1}$ too~\cite{balescu:75}. This
system (or hierarchy) of equations, the {\em BBGKY hierarchy\/}, can be
solved only if interrupted at some level. To break the hierarchy, say at
the $r$- th level, means expressing function $f_{r+1}$ with the help of
$f_r$ in equation for $f_r$. Such an interruption having been performed,
the equation for $f_r$ is in principle solvable. The solution may
subsequently be substituted in the equation for $f_{r-1}$. The procedure
may be continued until we are left over with the only equation for the
single- particle distribution function $f_1$. This is the {\em master\/}
({\em kinetic\/}) {\em equation\/}.

The mathematics leading to an $f_r$ independent of term $f_{r+1}$, can
obviously not be substantiated rigourously. One relies exclusively on more
or less acceptable heuristic semi-quantitative physical assumptions.

The first attempt in this direction was proposed in 1872 by L.~Boltzmann,
who derived an integro-differential equation for the one-particle
distribution function of a gas and proved that its solution proceeds  to
the equilibrium distribution function as time goes on. Among the
assumptions used, the most important one was the so-called molecular chaos
hypothesis, according to which particle momenta are independent of their
positions~\cite{Huang}. This hypothesis is eventually equivalent to the
assertion that the evolution of a system is not fully deterministic and
may be regarded as a stochastic process. Hence the derivation of second
law of thermodynamics (or H-theorem) from Boltzmann's equation cannot be
considered as the {\em proof\/} for approach of gases to equilibrium.
Despite this, Boltzmann's equation has become the foundation of the
kinetic theory of gases. The hypothesis of molecular chaos remained,
however, for more than a century the target of either criticism or efforts
to find its exact justification~\cite{Boltz_Eq}.

\subsection{Spectral Theory and Approach to Equilibrium}
\label{Appr}
A well-know theorem of functional analysis~\cite{Simon}, states that the
spectrum $\sigma (L)$ of any self-adjoint operator $L$, acting in a space
${\cal L}^2(\Gamma)$ of square-integrable functions, can be decomposed in
a unique way into three constituents: the pure point spectrum
$\sigma_{pp}(L)$ consisting of eigenvalues of $L$, the absolutely
continuous $\sigma_{ac}(L)$ and the singular spectrum $\sigma_{sing}(L)$.
Thus
\begin{equation}
\sigma(L) = \sigma_{pp}(L) \cup \sigma_{ac}(L) \cup \sigma_{sing}(L)\ .
\label{dec_sp}
\end{equation}

Let $\{ E_{\lambda}\}$ be the resolution of identity for operator $L$ and
$\nu_f = \langle E_{\lambda}f\,|\,f \rangle$ the spectral measure
generated by a function $f \in {\cal L}^2$. Decomposition~(\ref{dec_sp})
of the spectrum implies decomposition of ${\cal L}^2$ in orthogonal
subspaces
\begin{equation}
{\cal L}^2 = {\cal L}^2_{pp} \oplus {\cal L}^2_{ac} \oplus {\cal
L}^2_{sing}\ , \label{dec_L}
\end{equation}
where ${\cal L}^2_{pp} = \{ f \in {\cal L}^2\!:\!\nu_f$ is pure point\},
${\cal L}^2_{ac} = \{f \in {\cal L}^2\!:\!\nu_f$ is
absolutely continuous\} and ${\cal L}^2_{sing} = \{f \in {\cal
L}^2\!:\!\nu_f$ is singular\}~\cite{Simon}. The basis of ${\cal L}^2_{pp}$
consists of eigenfunctions of $L$. We shall denote $f \in {\cal L}^2_{pp}$
by $f_{pp}$, $f \in {\cal L}^2_{ac}$ by $f_{ac}$ and $f \in {\cal
L}^2_{sing}$ by $f_{sing}$. Decomposition~(\ref{dec_L}) means that every
$f \in {\cal L}^2$ can be written in the form

\begin{equation}
f = f_{pp} + f_{ac} + f_{sing}\ .
\label{decomp}
\end{equation}
It must be stressed that this decomposition is invariant under the action
of evolution operator $U_t$, i.~e.\ $U_t {\cal L}^2_{ac} = {\cal
L}^2_{ac}$, $U_t {\cal L}^2_{pp} = {\cal L}^2_{pp}$ and $U_t {\cal
L}^2_{sing} = {\cal L}^2_{sing}$. This means that each component of $f$
evolves under $U_t$ independently of the other ones. Hence, the
components evolve in orthogonal subspaces of ${\cal L}^2$. The dynamics of
the system can be, therefore, decomposed into subdynamics describing
qualitatively different types of evolutions.

In view of the above, any component of a distribution function evolves
with time independently of the other ones. One can, therefore, investigate
the
time evolution of each component separately. It turns out that for
realistic Hamiltonians (and therefore Liouvillians) $\sigma(L)_{sing} =
\emptyset$. Actually, the singular part of the spectrum corresponds to
``pathological'' Hamiltonians~\cite{Pearson}. For $f_{ac}$ it was shown
in~\cite{Kum84} that $U_t f_{ac}$ tends {\em weakly\/} to zero for $|t|
\to \infty$, i.~e.\ in this limit we have $\langle U_t f_{ac}\,|\,g\rangle
= 0$ for any $g \in {\cal L}^2$. In an alternative notation, $U_t f_{ac}
\stackrel{w}{\longrightarrow} 0$. Finally, the evolution of $U_t f_{pp}$
has been proved to be quasi-periodic.

Functions with the property $U_t f = f$ (for any $t$) span obviously a
linear subspace. Let us denote the latter by ${\cal L}^2_{const}$. For
each $f \in {\cal L}^2_{const}$ we get from~(\ref{Liouv}) $L f = 0$,
i.~e.\ $f \in \mbox{Ker} L$. The inverse implication is equally true, so
that we have ${\cal L}^2_{const} = \mbox{Ker} L$. A solution of equation
$U_t f \equiv f$ is an eigenfunction of operator $L$ corresponding to the
eigenvalue $\lambda = 0$. This yields $\mbox{Ker} L = {\cal
L}^2_{const}$~$\subseteq$~${\cal L}^2_{pp}$.

Let now $M \subset \Gamma$, $\mu(M) = \varepsilon$, $0 < \varepsilon <
\infty$ and let $\chi(M) \in {\cal L}^2(\Gamma)$ be the characteristic
function of the set $M$ (i.~e.\ $\chi(x)=1$ for $x \in M$ and $\chi(x)=0$
otherwise). For arbitrarily small $\varepsilon > 0$, the convergence
$U_tf_{ac} \stackrel{w}{\longrightarrow} 0$ implies
\begin{equation}
\lim_{|t| \to \infty} \langle U_t f_{ac} \,|\, \chi(M) \rangle = 0\ ,
\end{equation}
which is equivalent to
\begin{equation}
\lim_{|t| \to \infty} \int_M (U_t f_{ac})\, \d \mu = 0\ .
\label{fac}
\end{equation}
This resembles the behaviour of function $\sin (tx)$ on the finite
interval of $R$ and one can regard convergence~(\ref{fac}) as the
generalization of the Riemann-Lebesgue lemma to functions $f \in {\cal
L}^2_{ac}$.

The above implies that $U_t(f_{ac}+f_{const})$ tends weakly to
$f_{const}$. Weak convergence is mathematically a well defined property
with physically relevant contents. In physical systems, stress is rather
lead on the observables, or expressions like $\langle U_t f\,|\,g\rangle$,
where $g$ represents {\em dynamical\/} or {\em generating functions\/}.
Hence, if the Liouville operator acting in ${\cal L}^2(\Gamma)$ has an
absolutely continuous spectrum and if its only eigenvalue is $\lambda =
0$, then, for positively definite functions (distribution functions), as
$|t| \to \infty$, $\langle U_t f\,|\,g\rangle$ tends to $ \langle U_t
f\,|\,f_{const}\rangle $, that is the value of the observable
corresponding to an equilibrium state. This result may be regarded as the
exact criterion of approach to equilibrium in {\em conservative
systems\/}.

The above results show that approach to equilibrium in isolated systems
makes sense only for observables, not for distribution functions.
Distribution functions themselves change reversibly, and without any
apparent tendency to approach equilibrium. Observables related to
different distribution functions may tend to the same limit, meaning that
the approach to equilibrium is connected to a loss of information caused
by the integration leading to~$\langle U_t f\,|\,g\rangle $.

\subsection{The Brussels School Theory}
\label{Brussels}
As mentioned above, the spectral theory of operators shows
that the spectrum of any self-adjoint operator $L$ is real and that
operators defined as $U_t = e^{-iLt}$, with self-adjoint $L$, represent a
one-parameter group of unitary operators preserving the norm, i.~e.\
$\|U_t f\| = \|f\|$ for any $f \in {\cal L}^2$. Operators preserving the
norm are called isometric.

Unitary group $U_t$ is defined for any real $t$ and --- if such a group
describes the evolution of distribution functions --- this means, if an
evolution is possible for $t > 0$, then the evolution  for $t < 0$ is
equally possible. Hence, the unitary group describes a reversible
evolution, expressed explicitly by $U_{-t}(U_t f) = f$.

Irreversible evolution can be described only by semigroups (which cannot
be continued into groups), i.~e.\ sets of operators defined e.~g.\ only
for $t \geq 0$, not for $t < 0$. Let us denote such a semigroup by $W_t$.
The impossibility to define a semigroup for any real $t$ follows from the
spectral properties of its generator $K$:\ $W_t = e^{-iKt}$. If the
spectrum of $K$ contains points with negative imaginary components, the
semigroup is strictly contractive, i.~e.\ $\|W_t f\| < \|f\|$, so that it
is not isometric and it cannot be continued into a group~\cite{Kato}.

The reversible evolution is expressed by a group of isometric operators,
the irreversible one, on the contrary, by a semigroup of strictly
contractive operators. The most popular answer to the question of how such
a transition from a group to semigroup occurs when going from the
microscopic level to the macroscopic one is given by the Brussels school.
The short formulation of this viewpoint might run as follows. Suppose
that for a given system there exists a nonunitary operator $\Lambda$ such
that \\
(a) $W_t = \Lambda U_t \Lambda^{-1}$ is a strictly contractive semigroup
for $t \geq 0$, \\
(b) $W_t$ is positivity preserving, i.~e.\  $f(\omega) \geq 0$ for almost
all $\omega \in \Gamma$ implies $W_t f(\omega) \geq 0$ for almost all
$\omega$ too, and \\
(c) $W_t f_{const} = f_{const}$. \\
The dynamics $U_t$ of the system is then said to be {\em inherently
stochastic\/}, meaning roughly that the microscopic behaviour of the
system is deterministic, while it behaves as if it was really stochastic.
That is why it is appropriate to describe the dynamics using
statistical methods~\cite{Dougherty}. Under the above conditions it is
claimed specifically~\cite{MPC}, that operator $\Lambda$ converts the
deterministic evolution into a Markov process whose transition
probabilities can be obtained from the {\em adjoint\/} of $W_t$ by
\begin{equation}
P(\omega,\Delta;t)=W_t^{\ast} \chi(\Delta)\ ,
\label{Markov}
\end{equation}
where $P(\omega,\Delta;t)$ is the probability of transition  from point
$\omega$ to domain $\Delta$ of phase space in time $t$ and $\chi(\Delta)$
is a characteristic function of the set $\Delta$. Operator $\Lambda$ is
called the system's Lyapunov converter.

It is not expected that $\Lambda$ would exist for all dynamical systems
(flows). It is shown in~\cite{Misra} that the mixing property is necessary
and the condition of $K$-flow sufficient for the existence of a Lyapunov
converter $\Lambda$. The close connection between the intrinsic
irreversibility of the system, expressed by the existence of $\Lambda$,
and the instability of the motion is put forward.

The result of the Brussels school is important in that it divides the
systems into two categories:\ systems in which it makes sense to expect
thermodynamic behaviour and systems where this is not the case (trivial
examples of the latter are harmonic oscillators). This classification is
however used as the basis for derivation of far-reaching consequences
which are not always convincing.

The Brussels school does not give any physical interpretation to
the action of operator $\Lambda$. It is only stated that stochastic
evolution arises from a deterministic one ``simply as a result of `change
of representation' brought about by (non-unitary) similarity
transformation $\Lambda$''~\cite{MPC}. The physical meaning of this
transformation is indeed nowhere explained. Moreover, the analysis of the
properties of Markov processes shows that the claim, that every semigroup
$W_t$ of operators acting on ${\cal L}^2$ and having the properties
(a)--(c) (see above) defines a Markov process, is not in general
sufficient to
prove that the process has time-independent transition probabilities. This
condition is necessary for a process to have any physical meaning at all.

The action of $\Lambda$ is equivalent to a change of {\em dynamics\/} of
the system from deterministic to stochastic~\cite{Kum_Zakopane}. Such a
change can be performed in many ways and the operator $\Lambda$ is
constructed
precisely in such a way that the transition probabilities do not depend on
time. The possibility to change deterministic evolution into a stochastic
one does not give any information about the original deterministic
evolution, as is demonstrated by the following example~\cite{Kum87}.

Consider the unitary group of shifts $U_t f(x) = f(x+t)$, $f \in {\cal
L}^2(R; \d x)$. Using the operator of differentiation $\d /\d x$, this
group may be written, in the form
\begin{equation}
U_t = \e^{t\,\d /\d x}\ ,
\end{equation}
so that the generator of the group is operator $L = i\d /\d x$. For
this group it is possible to construct a Lyapunov converter $\Lambda$,
defined on the subspace of continuous functions with compact support,
converting $U_t$ into the strictly contractive semigroup $W_t = \exp
(t\,\d ^2/\d x^2)$, $t \geq 0$ associated with a one-dimensional diffusion
process, like diffusion of heat~\cite{Braunss}. The group of shifts $U_t
f(x) = f(x+t)$
lacks obviously any resemblance to a process that could be either random,
stochastic, approaching equilibrium, or even irreversible. The only
convergence property it has is the weak approach to a function $\varphi
\equiv 0$.

This example shows clearly that the existence of a Lyapunov converter (be
it that converting an isometric group into a {\em strictly\/} contractive
one) is not sufficient to prove the approach to equilibrium. Consequently
(in view of our comments about the relation between irreversibility and
approach to equilibrium) this cannot justify the origin of
irreversibility. However, it turns out that the notion of Lyapunov
converter is significant in understanding the essence of the method of
complex scaling introduced originally for identifying the so-called
resonant discrete points in the continuous spectrum of quantum
systems~\cite{Moiseyev,Brandas}.

\section{Quasi-isolated dynamical systems}
The discussion above was centred on the quest for solutions to
irreversible approach to equilibrium on the basis of reversible
microscopic laws in perfectly isolated systems. The proposed theories
leave us disappointed, being restricted to qualitative results and
concerned only with the limiting behaviour for $t \to \infty$. They are
unable to predict finite-time dynamics and experimentally verifiable
results.

Contrasting with the above let us not consider the system's boundaries
(walls) as a potential to be added to the Hamiltonian, but instead as the
locus for an independent contribution to the particles' motion. This is
the quasi-isolated system's paradigm.

\subsection{Experimental evidence}
Discussion of the directionality of time's arrow is often introduced
intuitively on the basis of a simplified representation of Gay-Lussac's
experiment. A box is considered, consisting of two compartments, the parts
being filled with gas at different pressures. Prior to the experiment the
gas is assumed to be at equilibrium. The long time evolution towards a new
equilibrium distribution following rupture of the division is taken to be
modelling irreversible behaviour of the global dynamics.
Joule repeated Gay-Lussac's experiment with great accuracy. His purpose
was to measure possible heat exchange with an external calorimeter
associated with spontaneous expansion. With an ideal gas, if no mechanical
work is allowed to be performed during the process, when the system had
reached its final state of equilibrium, no net exchange of heat with the
surroundings was observed. Joule concluded that the system behaved as if
it were isolated. Be it stressed that Joule was considering only the
initial and the final conditions, neglecting whatever dynamics was
involved in reaching final equilibrium.

Let us make the experiment more realistic by examining the effect of
pricking an air-inflated balloon inside either an acoustic reverberation
hall or an anechoic chamber. In both cases the excess air contained in
the balloon disseminates spontaneously throughout the rooms, never to come
back again, compressed in its initial volume, but the subsequent process
is very different indeed. In the reverberation hall an acoustic
perturbation is created and, the better the walls' reflecting quality, the
longer it remains. By contrast, in the anechoic room, the perturbation
vanishes promptly. In the reverberation room, some energy is stored in a
coherent or collective motion (acoustic perturbation) where it remains as
the memory of the initial conditions. With correctly shaped walls, the
initial information might even be partially retrieved as echoes. By
contrast, in the anechoic room, memory of the past is soon forgotten.

Initial and final conditions are identical in the two cases and so is the
air inside the rooms, and therefore the frequency and the quality of the
inter-particle collisions (Hamiltonian dynamics) assumed usually to be the
source of relaxation. The only difference between the two experiments is
the nature of the walls. One is therefore forced to conclude that global
relaxation dynamics of a spontaneously expanding gas leading to final
equilibrium depends on the acoustic (physical) quality of the walls
representing the
system's environment.

The experiment suggests convincingly that the global dynamics consists of
two independent major steps. According to the coupling efficiency of the
system to its surroundings (impedance matching), either step may be rate
determining. If coupling is very effective, global dynamics is controlled
by slow transport of mechanical properties to the walls (momentum, energy,
matter). Transport coefficients like viscosity, thermal conduction etc.,
are correctly defined only in such {\em non isolated\/} conditions. By
contrast, if the system is nearly isolated ({\em quasi-isolation\/}),
memory of the initial conditions remains for some time as a collective or
coherent motion of the particles (acoustic motion) and full thermodynamic
equilibrium is slow to reach. {\em Strict isolation\/} and transport
effects are incompatible.

The two steps involved by the scenario are very different in their
dynamics. Depending on the system of interest, they may be more or less
concomitant. For simplicity, we shall take them next as frankly separated
in the time.

In Joule's experiment, as soon as the membrane has been ruptured, a stream
of gas is ejected from the compartment at the highest pressure, thereby
creating a collective motion of the particles. By performing work on
itself, the system transfers energy into the jet. This is subtracted from
the initial thermal supply (adiabatic expansion). Loss of thermal energy
is equivalent to cooling.

On reaching the wall opposite the puncture, if this is hard, the initial
jet is reflected and turns progressively into a compound acoustic
perturbation with the same energy. The spectrum and phases of its
components are the memory of the initial conditions and of the shape of
the reverberating walls (coherence). This is the first step of the general
process. When this is done, although the particles are disseminated
throughout the whole volume, the system may not be claimed to be at
equilibrium ({\em weak irreversibility\/}).

Relaxation of the coherent or collective motion starts now. Every
collision with incoherently fluctuating wall atoms interrupts the running
canonical trajectory. A new one starts, with possibly modified initial
conditions\footnote{It should be obvious that this behaviour is much less
stochastic than the one put forward by Boltzmann's Sto{\ss}zahlansatz,
according to which stochasticity builds up at every collision between
two particles.}. Earlier correlations are progressively broken, thereby
thermalizing the energy accumulated initially in the jet and stored later
in the acoustic perturbation. As a result, if the system is an ideal gas
(hard spheres allowed), aforementioned transient cooling is progressively
neutralized, as is expected by Joule's result. When final equilibrium has
been reached, collective or coherent motion has relaxed and information
about the initial conditions is completely lost ({\em strong
irreversibility\/}).

Initial dissemination of the particles throughout the system follows
conservative Hamiltonian dynamics. No matter how intricate (chaotic) the
motion of individual particles may be (Sinai billiards), this part of the
motion preserves the memory of the initial conditions. Contrasting with
the latter, the correlations removing step, where the particular
properties of the walls determine how efficiently the system is coupled to
its surroundings introduces in the global motion stochastic non
Hamiltonian jumps between the different accessible trajectories. According
to whether during the particular impacts the relevant wall atom moves
towards the colliding particle or in the opposite direction, transient
work is transferred to the system or to the environment. Energy fluctuates
about its average value. Only if the thermodynamic requirement is
fulfilled that the system and its neighbourhood are at the same
temperature does the average energy transfer vanish.

\subsection{Thermodynamics}
\subsubsection{Entropy}
Any function determined completely by the set of constraints defining the
relevant system's particular macrostate is a function of state. In 1865
Clausius discovered a function of state called entropy which, for
reversible processes, is defined as a differential $\delta S = \delta
Q/T$, where $T$  is the system's temperature. In 1877, Boltzmann derived
an expression that links the experimental entropy to the statistical
properties of the relevant macroscopic system. This reads
\begin{equation}
S=k_B \ln [W(A)],
\label{entropy}
\end{equation}
with  $W(A)$ meaning the probability associated to macrostate $A$. The
latter is to be interpreted as the total volume accessible to the
motion in phase space, given the set of constraints (represented by the
collective variable $A$) describing the system's particular macrostate.
Let it be noted that equilibrium macrostates are usually defined by their
total energy $E$, particle number of any sort $N_r$ and physical volume
$V$, the traditional microcanonical variables. In the literature,
extension of the discussion to non-equilibrium macrostates is avoided.
This limitation will be reconsidered below.

In a strictly conservative isolated environment, the dynamics being
described by a single multi-particle trajectory in phase space, no matter
how intricate (chaotic) this may be, transitions between different
trajectories are not possible. Then, according to the definition, the
entropy is zero and it does never change. This conclusion is consistent
with Liouville's theorem claiming conservation of the measure in phase
space when the mechanics is conservative.

For Boltzmann's entropy to be a pertinent function of state, prompt
accessibility of all the quantum states or trajectories belonging to the
given macrostate is required. Accessibility means incoherent
transitions between the available and accessible trajectories or quantum
states during the observation period. This depends on fast uncorrelated
action of the environment with fluctuating exchange of mechanical
properties (momentum, energy). As a corollary, and as expected by the
statistical nature of the thermodynamic functions, it appears that the
definition of the entropy implies some averaging over the time. The
resolution linked to the definition of the entropy is the average lifetime
of conservative trajectories. With macroscopic systems, where the impact
rate with boundaries goes to infinity, the average lifetime and hence the
time resolution tend to zero.

Relaxation implies relief of constraints. It opens the way to an enhanced
choice of trajectories (microstates). Accessibility of more trajectories
increases Boltzmann's entropy.

In describing equilibrium states, the extensive variables mentioned
traditionally are the basic microcanonical constraints $E$, $V$ and $N_r$.
In order to specify unambiguously non-equilibrium macrostates, where more
constraints prevail, additional extensive properties must be included.
This may be for example the momentum associated with a possible collective
or coherent motion of the system, where some of the total energy is stored
(e.g.\ the jet or the acoustic motion in the aforementioned Gay-Lussac
experiment). Many other possible distortions with respect to equilibrium
may occur, like moments of the energy or density distribution, etc.

Let the list of the extensive properties defining the constraints of a
macroscopic system in a particular macrostate be written $\{X_l\}$. The
entropy is a function of this collection of variables. By differentiating
the entropy with respect to the set we get by definition the set of
conjugate intensive variables or intensities $\{\xi_l\}$:
\begin{equation}
\d S= \sum_l\, \frac{\partial S}{\partial X_l}\,\d X_l
=-k_B\, \sum_l \,\xi_l\,\d X_l.
\label{Gibbs}
\end{equation}
This equation may be considered as defining the temperature $(\partial
S/\partial E)^{- 1}$ and the chemical potential $-T(\partial S/\partial
N_r)$. In non-equilibrium conditions it generalizes all the definitions by
proposing an intensity conjugate to each of the additional non-equilibrium
constraints.

Equation~(\ref{Gibbs}) is Gibbs' celebrated equation, generalized to
non-equilibrium ma\-cro\-states. In the simplified model of a
spontaneously expanding jet mentioned above (velocity of the collective
motion is ${\vec v}$), the new version of Gibbs' equation reads
\begin{equation}
\d S=\frac{\d E}{T}+{{\cal P}\over {T}}\d V-\sum_r\frac{\mu_r}{T}\d N_r-
k_B\,{\vec{\sigma}} \cdot \d {\vec P},
\label{jet}
\end{equation}
where ${\vec P}=Nm{\vec v}$ represents the collective momentum of the jet
and ${\vec \sigma}$ the conjugate intensity. It may be shown~\cite{xh:92}
that ${\vec \sigma}={\vec v}/k_BT$. In the last term of
equation~(\ref{jet}), the differential of the collective or coherent
energy is easily recognized. We have therefore equivalently
\begin{equation}
\d S=\frac{\d E}{T}+{{\cal P}\over {T}}\d V-\sum_r\frac{\mu_r}{T}\d N_r -
\frac{1}{T}\d \,(coherent\; energy).
\label{coherent}
\end{equation}

Energy conservation throughout the expansion makes $\d E=0$ and isolation
causes $\d N_r=0$. During the adiabatic dissemination period, the second
term (work made available by expansion) is very exactly balanced by the
last contribution (energy stored in the coherent motion), making $\d S=0$,
in agreement with Liouville's theorem for isolated conservative motions.
Final relaxation involves transformation of the coherent motion into
thermal energy. When this has been achieved, thanks to stochastic exchange
at every impact with the boundaries, the integral of the last term
vanishes and Gibbs' equation yields the correct final equilibrium entropy
after expansion.

If the contribution regarding the non-equilibrium constraint had been
omitted in equations~(\ref{jet}) and~(\ref{coherent}), we would not have
been able to describe the thermodynamics of the low entropy
non-equilibrium transient state.

\subsubsection{The generalized Massieu function}
The inconvenience of considering the entropy as the leading thermodynamic
function is that it is an explicit function of the extensive properties
($X_l$), while the intensities ($\xi_l$) are better measured and
controlled by the environment. That is why thermodynamics makes widely use
of potentials and other Massieu-Planck functions, obtained from the
entropy or the energy by Legendre transformations.

Most popular are free energy transformations. However, contrasting with
the second law concerning the entropy, general laws involving the energy
do not exist. It is therefore advisable to consider transformations
involving the entropy itself. If all the parameters (excepting the
system's physical volume $V$) defining non-equilibrium conditions are
included in the transformation, we obtain the generalized Massieu function
${\cal M}(\xi_l,V)$:
\begin{equation}
{\cal M}(\xi_l,V)=\frac{S}{k_B}+\sum_l\xi_l X_l.
\label{Massieu}
\end{equation}
Unlike Massieu's original proposal, ${\cal M}$ is an explicit function of
all the state defining intensities. It may be verified that
\begin{equation}
\frac{\partial {\cal M}(\xi_l,V)}{\partial \xi_j}=X_j.
\label{extensive}
\end{equation}

The advantage of referring to a state function depending explicitly on
intensities is that, with promptly exchangeable properties, the relevant
intensities of the system of interest remain at all times equal to their
values in the neighbourhood. We might call them strong intensities (e.g.
the temperature in an efficiently thermostated system). Dynamics of
transient states refers to the intensities of the rate determining slowly
exchanging or soft properties.

The Legendre transformation changes the maximum entropy condition with
respect to fluctuations of the extensive variables into a minimum of the
generalized Massieu function with respect to the intensities relating to
non-ex\-change\-able properties (e.g. particle numbers and their
distribution in closed systems). With transient effects, this fundamental
property defines the path followed by the system during relaxation. It
gives a key for treating coupled flows.

The equations above (\ref{Massieu}--\ref{extensive}) are generally valid.
With ideal gases, the expression for the generalized Massieu function
takes a very simple form. Individual motions being independent, the global
motion may be represented by a swarm of points in a reduced
6-dimensional single-particle phase space ($\Gamma_1$). From here on,
$\Gamma$ will represent a one-particle phase space.

Let $f(x)$, $x \in \Gamma$, be the most probable particle distribution:
that which
maximizes the entropy. Any extensive property $X_j$ is then related to a
generating function $\phi_j(x)$ so that
\begin{equation}
X_j=\int_{\Gamma} \phi_j(x) f(x) \d \Gamma \equiv \langle \phi_j | f
\rangle \, .
\label{idealextensive}
\end{equation}
In that context, function $f(x)$ is readily known to be~\cite{xh:92}
\begin{equation}
f(x)= \exp [\sum _l\xi_l \phi_l(x)],
\label{dis_fun}
\end{equation}
where the intensities $\{\xi_l\}$ are the Lagrange multipliers used in the
maximizing process for the entropy.

With the latter distribution function, it may be verified that ${\cal M}$
takes the very simple form
\begin{equation}
{\cal M}(\xi_l,V)= \int_{\Gamma} \exp [ \sum
_l\xi_l\phi_l(x)] \d \Gamma.
\label{idealMassieu}
\end{equation}
Its numerical value is the (average) number of particles contained in the
system. Through the integration limits in configuration space it has the
system's physical dimensions ($V$) as one of its independent variables. By
restricting the integration to the only momentum coordinates, a local
generalized Massieu function is obtained, the value of which represents
the average local density in configuration space.

With real gases, the generalized Massieu function is modified due to the
interaction potential between the particles. The simplified formulation is
however still useful as a low density approximation when the duration of
the inter-particle collisions is negligible compared to the time
separating the collisions (e.g. hard spheres).

\section{Transport coefficients}
The main objective of the theory is to predict transport coefficients from
first principles, to be compared with the experiment. Since Boltzmann's
equation there has been a considerable literature concerning that
question~\cite{balescu:75,clarke:76}. Most frequently cited are the
traditional Chapman and Enskog derivations~\cite{hirsch:54} and the more
recent Green-Kubo formalism~\cite{balescu:75}.

When referring relaxing systems to their fixed boundaries, the
thermodynamic approach fits best into an Eulerian
frame~\cite{ryhm:85,goldstein:51}:
\begin{equation}
\label{liouville2}
\frac{\d f_N}{\d t}=\{f_N,H\}+J.
\end{equation}
Poisson bracket $\{f_N,H\}$ expresses implicit deterministic contribution
to the motion while the source/sink term $J$ describes explicit stochastic
action of the environment. If $\d f_N/\d t\neq 0$, we have a transient
state. Either implicit or explicit contributions may then be rate
determining. If $\d f_N/\d t= 0$, $J$ may still be different from zero,
balancing a non-zero $\{f_N,H\}$. That are stationary states. Steady
transport of heat between reservoirs at different temperatures and steady
transport of momentum in the Couette flow belong to that class of
processes. Stochastic action of the neighbourhood defines a finite
life-time to deterministic trajectories. In stationary conditions, flows
of
extensive properties supported by $J$ are obtained by integrating the
deterministic contribution over this average life-time. For individual
particles, this is the average periodicity $\tau$ of effective relaxing
collisions.

In very low density systems, when the mean free path is comparable to the
system's physical dimensions (Knudsen gas), properties picked up by the
particle at one wall are transported in a single jump to the opposite one.
In the thermodynamic limit (non-Knudsen regime), head-on collisions of
identical particles do not hamper transport properties but parallactic or
off-axis inter-particle collisions do. They reduce the range of free
transport, while information about the reservoir conditions is transferred
to the relevant bulk region.

We consider a given extensive property $X_j$ with generating function
$\phi_j(x)$, $x \in \Gamma$, and we investigate its flow along the $z$-
direction. Let
$z^{*}$ be the ordinate of an arbitrary plane. The basic equation for the
relevant flow $J_j$ through this plane is
\begin{equation}
\label{flow}
J_j=\frac{1}{\tau}\;\int \int \int(\frac{\d ^3{\vec p}}{h^3})
\int_{(z^{*}-p_z \tau/m)}^{z^*} \, \phi_j(x) \, \exp[ \sum _l\xi_l
\phi_l(x)] \, \d z.
\end{equation}
The term $1/h^3$ is introduced for the sake of
normalization~\cite{pathria:72}.

Equation~(\ref{flow}) stresses that free transport of the given property
by particles is limited to the life-time of their trajectories. Local
thermodynamic conditions at the latter's onset determine how much of the
property is transported. Hence, the lesser the collision frequency, the
more effective is the transport. Collisions increase the system's
resistance to flow.

When integrated, the effective collision frequency comes in the normalized
form $\tau/(D\sqrt{\beta m})$. With Knudsen systems (mean free path at
least of the order of the system's physical dimensions), this parameter
equals 1. In thermodynamic conditions, the parameter is much less than 1,
justifying expanding the integrand to its lowest order in $\tau$.

In the following, equation~(\ref{flow}) will be applied to different types
of flows.

\subsection{Single-component gases}
\subsubsection{Viscosity}
We consider a fluid bound by a pair of walls distant by $2D$, moving in
opposite directions (Couette flow). The system's stationary conditions are
defined completely by the set of constraints listed in
table~\ref{gen_viscosity}. The intensities under direct control of the
surroundings are the particles number, the kinetic energy and the
intensity conjugate to the gradient of shear momentum. It may be
verified~\cite{xh:92} that the velocity of the walls ($y-$direction)
equals $\pm \sigma_y/\beta$.

\begin{table}
\caption{Constraints for Couette flow ($\zeta=z/D$)}
\label{gen_viscosity}
\begin{tabular}{ccc}
\hline
$X_l$ & $\phi_l(\Gamma)$ & $\xi_l$ \\
\hline
Particles number & 1 & $\alpha$  \\
2nd moment of particle distribution & $[\zeta^2-1]$ & $\theta_2$ \\
Kinetic energy & $\sum({p^2/{2m}})$ & $-\beta$ \\
2nd moment of energy distribution & $(\zeta ^2-1)\,\sum({{p^2}/{2m}})$ &
$-\gamma_2$ \\
Gradient of shear momentum & $\zeta\, p_y$ & $\sigma_y$ \\
\hline
\end{tabular}
\end{table}

Two variables remain to be determined, namely $\theta_2$ and $\gamma_2$,
requiring two independent equations. In stationary conditions, pressure
gradients or acoustic perturbation are absent. This is expressed by
vanishing $z^*$-dependence of flow of transverse momentum ($p_z$).
Likewise, the total flow of energy through the system is zero. By
implementing equation~(\ref{flow}) with the two relevant generating
functions, the conditions $\partial J_{p_z} / \partial z =0$ and $J_E=0$
yield together
\begin{equation}
\label{theta_gamma}
\theta_2=0,\qquad \qquad \qquad \frac{5}{2} \; \frac{\gamma_2}{\beta}
=\frac{m\sigma^2}{2\beta}.
\end{equation}

Flow of the shear component of momentum ($p_y$) is obtained by
implementing equation~(\ref{flow}) with generating function $p_y$, where
$\theta_2$ and $\gamma_2$ are replaced by their values. This yields
\begin{equation}
\label{momentum_flow}
J_{p_y}=-\frac{\sigma_y}{2\beta D}\;\frac{n \tau}{\beta },
\end{equation}
where $n={\cal M}/V$ is the particle density.

The coefficient of shear viscosity is the ratio of the forces applied to
the plates, compensating for transfer of momentum from wall to wall, to
the velocity gradient ($\sigma_y/(\beta D)$). Following
equation~(\ref{momentum_flow}), its value is
\begin{equation}
\label{viscosity}
\eta=n\; \frac{\tau}{\beta}.
\end{equation}

\subsubsection{Thermal conduction}
Let us consider now a system in thermal contact with two planar heat
reservoirs at different temperatures separated by $2D$. The system's
stationary conditions are completely described by the set of constraints
listed in table~\ref{gen_conduction}. By inspecting the generating
function conjugate to the temperature gradient it is clear that $k_B\nabla
T =-\gamma_1/(\beta^2D)$.

In the presence of a temperature gradient, particles moving towards the
cold wall have been equilibrated with the system upstream in a hotter
region at the instant of their last collision and vice-versa. Hence, in
moving from the hot wall to the cold one, particles travel on the average
faster than in their return cycle. If the particles are to change their
average kinetic energy in a correlated fashion on impact with either
walls, while the container (the pair of walls) is to remain on the average
immobile, collective momentum must be transferred by the container into
the system. That is why a generating function for collective motion of the
particles perpendicularly to the walls needs to be considered in
constructing the expression for flow of heat.

\begin{table}
\caption{Constraints for thermal conduction ($\zeta=z/D$)}
\label{gen_conduction}
\begin{tabular}{ccc}
\hline
$X_l$ & $\phi_l(\Gamma)$ & $\xi_l$ \\
\hline
Particles number & 1 & $\alpha$  \\
Gradient of particle distribution & $\zeta$ & $\theta_1$ \\
Kinetic energy & $\sum({p^2/{2m}})$ & $-\beta$ \\
Gradient of energy distribution & $\zeta/D\;\sum({{p^2}/{2m}})$ &
$-\gamma_1$ \\
Collective momentum & $ p_z $ & $\sigma_z$ \\
\hline
\end{tabular}
\end{table}

The intensities under direct control of the surroundings are the particles
number, the average kinetic energy and the temperature gradient. Two
intensities remain to be determined: $\theta_1$ and $\sigma_z$. This
requires two independent equations. One is stationarity. The other
equation describes mechanical equilibrium of the system between its walls.

Let $z^*$ be an arbitrary position between the boundaries. The average
local density $n$ responds to the equation
\begin{equation}
\label{density}
n(z^*)=\frac{1}{h^3}\int \int \int \d p_x  \d p_y  \d p_z \exp\{ \sum
_l\xi_l \,\phi_l [(z=z^*),p_x,p_y,p_z]\}.
\end{equation}

We call $n_+(z^*)$ the partial density of the particles with positive
velocity along the $z-$direction. Stationarity implies that this partial
density equals the sum of the densities of the particles present at places
from where they will be reaching this position without disturbance after
one collision period, their velocities being oppositely oriented. Hence
\begin{eqnarray}
\label{n+}
n_+(z^*) =  \frac{1}{h^3}\int_{-\infty}^{\infty} \d p_x \int_{-
\infty}^{\infty} \d p_y \int_{-\infty}^0 \d p_z \qquad \qquad \qquad
\qquad
\nonumber \\
\exp \{ \sum _l\xi_l\, \phi_l[(z=z^* - \frac{p_z\tau}{m}),
p_x,p_y,p_z]\}.
\end{eqnarray}
This condition yields
\begin{equation}
\label{sigma}
(\theta_1-2 \frac{\gamma_1}{\beta})\;\frac{\tau}{mD}=2\sigma_z.
\end{equation}

The second condition expresses position independence of flow of momentum
across the system. Equation~(\ref{flow}) is used with $p_z$ as the flow
defining generating function. Condition $\partial J_{p_z}/\partial z=0$
yields
\begin{equation}
\label{mecha}
\theta_1=\frac{5}{2}\;\frac{\gamma_1}{\beta}.
\end{equation}

For flow of energy (heat) through the system, the generating function
in equation~(\ref{flow}) is $\sum(p^2/2m)$. If internal rotation
is superimposed on translation (Eucken's correction
\cite{hirsch:54,eucken:13}), the relevant contribution should be added to
the generating function. With atomic gases however, the result reads
\begin{equation}
\label{heatflow}
J_E=\frac{15}{8}\;\frac{\gamma_1}{\beta^2D}\;n\;\frac{\tau}{\beta m}.
\end{equation}

Heat conductivity ($\lambda$) is the ratio between the sum of the
exchanges at either walls ($2J_E$) and the temperature gradient. Hence
\begin{equation}
\label{conductivity}
\lambda=\frac{15}{4}\;k_B\;n\;\frac{\tau}{\beta m}.
\end{equation}

In equations~(\ref{viscosity}) and~(\ref{conductivity}), the transport
coefficients are expressed in terms of the effective collision periodicity
$\tau$. For absolute comparison with the experiment, an additional
expression is required that relates the latter to the mechanical
properties of the colliding species (mass and cross-section) at the given
temperature. Without this additional information, only the ratio between
viscosity and heat conductivity may be compared with experimental data.
This ratio is by definition Prandtl's number
\begin{equation}
\label{prandtl}
{\rm Pr}=\frac{\eta \, c_p}{m\,\lambda},
\end{equation}
with $c_p$ as the constant pressure heat capacity. Implementation with the
results obtained above yields the experimental results identically.

\subsection{Mixtures of atomic gases}
Let the components of a given mixture be indexed $A$ and $B$, where $A$
points to the substance with the higher mass. Each component may be
considered as a separate subsystem, with its own thermodynamic variables,
interacting simultaneously with the other one and with the environment.
The generalized Massieu function being extensive, we have for the
composite system
\begin{equation}
\label{massieu_AB}
{\cal M}={\cal M}_A + {\cal M}_B.
\end{equation}

With dilute gases or gases interacting as hard spheres, the individual
generalized Massieu functions are defined as in
equation~(\ref{idealMassieu}). For each component separately the
generating functions to be used are the same as for single-component gases
(see tables~\ref{gen_viscosity} and~\ref{gen_conduction}), excepting for
the requirement of indexing the relevant masses in the appropriate
generating functions. In stationary or quasi-stationary conditions (see
below), for exchangeable properties where equilibrium between the
subsystems prevails, the intensities are the same. In the examples treated
below, that is the case for the temperature and its moments and for the
intensities conjugate to collective motion. Intensities conjugate to not
exchangeable properties and their respective moments will be indexed
according to the particular component they refer to.

The collision periodicity has been shown above to be an essential
ingredient in the dynamics. In multi-component systems, there is an
average collision periodicity for each of the constituents ($\tau_A,
\tau_B$). It measures for each how long the relevant atoms move freely
before being halted by the matrix formed by the remaining particles.

In multi-component systems there are homogeneous and heterogeneous
collisions. Their frequencies add up. The efficiency for exchange of
momentum from a colliding atom to the local thermodynamic bath depends on
the masses of the collision partners. When a heavy particle hits a light
constituent of the thermodynamic bath, its path is less disturbed and less
momentum is transferred than in the opposite case.

We assume a particle with mass $m_1$ and linear momentum ${\vec P}$
hitting a stationary matrix particle with mass $m_2$. If the exit path of
the matrix particle forms an angle $\psi$ with the incident one, momentum
transferred to the matrix equals $2|{\vec P}|\cos(\psi)m_2/(m_1+m_2)$.
Hence, the relative transfer efficiency of heterogeneous collisions is
$2m_2/(m_1+m_2)$. For the total effective collision frequency of atoms of
one sort with respect to the matrix ($1/\tau$), the latter coefficient is
the appropriate scaling factor relating efficiency of heterogeneous
collisions to homogeneous ones.

With hard spheres, the traditional expression for the collision
periodicity between identical particles with collisional cross-section $d$
is known to be~\cite{hirsch:54}
\begin{equation}
\label{periodicity}
\tau = \frac{5}{16}\;\frac{\sqrt{\pi m \beta}}{n \pi d^2}.
\end{equation}
Let now the collision cross-sections be respectively $d_{AA}, d_{BB},
d_{AB}$. Using the efficiency parameter defined above and adding for
either constituents the homogeneous and the heterogeneous contributions,
the total effective collision periodicities read
\begin{eqnarray}
\label{periodicity_AB}
\tau_A = \frac{5}{16}\;\sqrt{\frac{\beta}{\pi}}\;\left( \frac{n_A
d_{AA}^2}{\sqrt{m_A}} + \frac{2m_B}{m_A+m_B}{n_B d_{AB}^2}
\sqrt{\frac{m_A+m_B}{2m_Am_B}} \right)^{-1}, \\
\tau_B = \frac{5}{16}\;\sqrt{\frac{\beta}{\pi}}\;
\left(\frac{2m_A}{m_A+m_B} n_A d_{AB}^2 \sqrt{\frac{m_A+m_B}{2m_Am_B}}+
\frac{n_B d_{BB}^2}{\sqrt{m_B}}\right)^{-1}.
\end{eqnarray}
From here on it is advisable to replace $n_A$ by $xn$ and $n_B$ by $(1-
x)n$.

With hard spheres we have $d_{AB}=(d_{AA}+d_{BB})/2$. It appears that the
published data on the viscosity of mixtures of atomic gases is accurate
enough to allow the heterogeneous hard sphere cross-section to be
corrected by a factor $\epsilon$ close to 1.

\subsubsection{Viscosity}
With Couette flow conditions, the intensities under direct control of the
surroundings are the intensities conjugate to the particle numbers of
either substances ($\alpha_A,\, \alpha_B$), the temperature (or better
$\beta$) and the intensity conjugate to the linear moment of shear
velocity $\sigma_y$ (see table \ref{gen_viscosity}). Three intensities
need still to be determined, namely the quadratic moment of the
temperature (or better $\gamma_2$) and the quadratic moments of the
particle distributions for $A$ and $B$ ($\theta_{2,A}, \,\theta_{2,B}$).

The three additional relations required for completing the thermodynamic
description of the system are of the same vein as those used for Couette
flow in single component gases. For symmetry reasons, it is easy to show
that flow of shear momentum is independent of the particular values of
three missing intensities. The principles involved in their determination
will be therefore skipped.

Flow of momentum is supported by either components. For each, the
contribution is given according to equation~(\ref{flow}), where the
generating function to be implemented as $\phi_j$ is $p_y$. Integration
yields
\begin{equation}
\label{visc_flowAB}
J_{p_y}=-n\frac{\sigma_y}{2\beta^2 D}\;[x \tau_A + (1-x) \tau_B].
\end{equation}
The viscosity of the mixture is therefore
\begin{equation}
\label{visc_AB}
\eta_{mix}=\frac{n}{\beta}\;[x \tau_A + (1-x) \tau_B],
\end{equation}
where $n/\beta$ is the total pressure.

The result of equation~(\ref{visc_AB}) is plotted in figure~\ref{vishexe}
for a  mixture of Xe in He. Experimental results at 291~K
~\cite{toulouvis:75} are indicated on the same graph (experimental
uncertainties $\approx \pm 1\%$). Correction factor $\epsilon$ for
heterogeneous collisions may be estimated by fitting the curve to the
experimental results. That obtained without the correction factor is
displayed as a dotted curve. The fit performed on the ten mixtures of
atomic gases leads to the values of $\epsilon$ in the range 1.03 -- 0.98,
the highest values being for mixtures with a light component (He). Results
for the ten mixtures of atomic gases are listed in~\cite{Hens}.
\begin{figure}
\vspace{10cm}
\caption{Predicted and experimental viscosity of a mixture of Xe in He at
271~K. The smooth curve is for $\epsilon=0.98$, the dotted curve for
$\epsilon=1$.}
\label{vishexe}
\end{figure}

\subsubsection{Diffusion and thermal conduction}
For a binary system enclosed between two reservoirs at different
temperatures separated by a distance $2D$ ($x$: mole fraction of the
heavier substance ($A$)), stationary conditions are completely described
by the set of constraints listed in table\ref{gen_conduction}, the
intensities being indexed accordingly.

The intensities conjugate to the particle numbers of either substances
($\alpha_A,\, \alpha_B$), the temperature (or better $\beta$) and its
gradient (or better $\gamma_1$) are under direct control of the
surroundings. Thermal interaction between the subsystems removes the
necessity for indexing the latter two intensities.

Three intensities remain to be determined, namely the two gradients of the
particle distributions ($\theta_{1,A}, \,\theta_{1,B}$) and the intensity
conjugate to collective momentum from wall to wall ($\sigma_z$). Hence,
three additional conditions or equations are required. Two are identical
to the conditions discussed for single component systems: mechanical
equilibrium and global stationarity.

Mechanical equilibrium of the system between its walls implies vanishing
total pressure gradient. It does not require {\em per se\/} vanishing
partial pressure gradient for either substances separately. A possible
pressure gradient of $A$ is indeed neutralized by an opposite gradient for
$B$. By stating that the sum of the contributions of either substances to
flow of momentum between the boundaries is position independent, the
following equation is derived (compare equation~(\ref{mecha}))
\begin{equation}
x \left(\theta_{1,A} - \frac{5}{2}\frac{\gamma_1}{\beta}\right)
+ (1-x) \left(\theta_{1,B} - \frac{5}{2}\frac{\gamma_1}{\beta}\right) =0.
\label{mecha_AB}
\end{equation}

The condition for global stationarity is defined along the same lines as
above (equations~(\ref{density})--(\ref{sigma})), where the density
$n_+(z^*)$ is now understood as the sum for the two components. As a
result, the relation for internal collective motion ($\sigma_z$) becomes
(see equation~(\ref{sigma}))
\begin{eqnarray}
\label{sigma_AB}
x\sqrt{m_A}\left[\sigma_z - \frac{1}{2} \left(\theta_{1,A}-2
\frac{\gamma_1}{\beta}\right)\;\frac{\tau_A}{m_AD}\right] \qquad \qquad
\qquad \qquad \nonumber \qquad \qquad \\
+ (1-x)\sqrt{m_B}\left[ \sigma_z - \frac{1}{2}
\left(\theta_{1,B}-2
\frac{\gamma_1}{\beta}\right) \; \frac{\tau_B}{m_BD} \right] = 0.
\end{eqnarray}

The last condition to be considered concerns mutual diffusion or motion of
the subsystems with respect to each other. By implementing
equation~(\ref{flow}) with the generating function $\phi_j=1$, the
particle flow of either subsystems is obtained. According to whether the
named parameters are indexed $A$ or $B$, the results are
\begin{equation}
\label{particleflow}
\begin{array}{l}
J_A  =  \frac{xn}{\beta}\left[\sigma_z - \frac{1}{2}\left(
\theta_{1,A}-\frac{5}{2}\frac{\gamma_1}{\beta}\right)\frac{\tau_A}{m_A
D}\right],\\
J_B = \frac{(1-x)n}{\beta}\left[\sigma_z - \frac{1}{2}\left(
\theta_{1,B}-\frac{5}{2}\frac{\gamma_1}{\beta}\right)\frac{\tau_B}{m_B D}
\right].
\end{array}
\end{equation}

The first contribution in either equations (that proportional to
$\sigma_z$) represents collective drag generated in the fluid by
correlated effect of the walls. This acts on the two subsystems alike.
Therefore it does not drive diffusion of the subsystems with respect to
each other. By contrast, diffusion is related to the second part of the
flow equations. As it may be verified, this is driven by the relevant
partial pressure gradients.

Diffusive stationarity is reached by differential displacement of the
subsystems with respect to each other. Then we have for either subsystems
vanishing partial pressure gradients. This is the remaining constraint for
complete thermodynamic description of the system. In experimental
conditions, the question is however whether diffusive stationarity has
been reached in practical cases when thermal conductivity of
multi-component mixtures is measured.

For measuring thermal conductivity, an appropriate binary mixture is
prepared in a conventional thermostat. The walls are then brought at
different temperatures. When possible acoustic perturbations have relaxed,
the total pressure distribution is flat (equation~(\ref{mecha_AB})).
Nevertheless, on establishing the temperature gradient, pressure gradients
of the individual constituents are created, forcing the
particles to segregate. If the mixture consists of particles with
different mobility, final stationary conditions are slow to reach. The
slower moving particles tend to remain distributed homogeneously, as
they were before the temperature gradient was created. The partial
pressure gradient of the faster moving subsystem compensates for resulting
unbalance. Final equilibrium requires that the migration of the slower
particles have taken place, cancelling all partial pressure gradients.

The published experimental data that have been considered do not mention
whether (or how much) the system has been allowed to relax the initially
created individual partial pressure gradients.
Let us assume this would not have occurred at all. The state is then
pseudo-stationary as it continues to change slowly in time while the
constituents still migrate with respect to each other. The two
subsystems should then be considered as acting independently for all the
properties concerning the particle distributions. They remain
however tightly coupled for the properties that are promptly interchanged.
In particular, they share the same $\beta$ and $\gamma_1$. The intensity
$\sigma_z$ conjugate to the collective momentum generated by the
temperature gradient is also common to the two subsystems. Concerning the
latter, its relation to the other intensities and to the collision
periodicities is given by (\ref{sigma}). Instead of (\ref{sigma_AB}) we
have now two relations, namely
\begin{equation}
\begin{array}{l}
\left(\theta_{1,A}-2 \frac{\gamma_1}{\beta}\right)\;\frac{\tau_A}{m_A D}
= 2\sigma_z, \\
\left(\theta_{1,B}-2 \frac{\gamma_1}{\beta}\right)\;\frac{\tau_B}{m_B D}
= 2\sigma_z.
\end{array}
\end{equation}

By combining (\ref{mecha_AB}) with the two latter ones, an expression for
the gradients of the individual partial pressures may be derived. Writing
\begin{equation}
\label{R}
R=\frac{\tau_A/m_A}{\tau_B/m_B},
\end{equation}
this relation reads
\begin{equation}
\label{partial_pressure}
\theta_{1,A}-\frac{5}{2}\frac{\gamma_1}{\beta}=\frac{1}{2}\,
\frac{(1-R)(1-x)}{(1-x)R +x}\, \frac{\gamma_1}{\beta}.
\end{equation}

In practical cases, depending on the mixture to be considered, when heat
conductivity is measured, the system may be somewhere between the two
extreme conditions. The uncertainty concerning how close diffusion has
reached stationarity in the experimental conditions where the measurements
have been performed, explains why thermal conduction data of mixtures are
difficult to reproduce. Let us express the uncertainty by a coefficient
$c$, to multiply the right-hand side of equation~(\ref{partial_pressure}).
When discussing a homogeneous set of data with varying compositions $x$,
we assume for simplicity that the same coefficient is valid. Equilibrium
for diffusion implies $c = 0$.

\begin{figure}
\vspace{10cm}
\caption{Thermal conductivity of a mixture of Xe in He at 271~K. The
smooth curve is for $c=0.5$}
\label{conhexe}
\end{figure}

Transport of heat is supported by either components of the mixture. For
each, the contribution is given according to equation~(\ref{heatflow}),
where the relevant intensities are determined as above. Hence,
\begin{eqnarray}
\label{heatflow_AB}
J_E  =  \frac{xn}{\beta^2}\left[\sigma_z - \frac{1}{2}\left(
\theta_{1,A}-\frac{7}{2}\frac{\gamma_1}{\beta}\right)\frac{\tau_A}{m_A D}
\right] \qquad \qquad \qquad \qquad \qquad \nonumber \\
+ \frac{(1-x)n}{\beta^2}\left[\sigma_z - \frac{1}{2}\left(
\theta_{1,B}-\frac{7}{2} \frac{\gamma_1}{\beta}\right)\frac{\tau_B}{m_B D}
\right].
\end{eqnarray}

In comparing the result with experimental data, coefficient $c$ may be
taken as an adjustable parameter. Using the data published by E. Thornton
and coworkers~\cite{touloucon:70} for 291~K, coefficient $c$ has been
found to range between 0 for light--light  mixtures (e.g. He--Ne) and 1
for heavy--heavy mixtures (Kr--Xe). Figure~\ref{conhexe} is an
illustration of the results for the He--Xe mixture, where the value of $c$
optimizes at 0.5. Results for the ten mixtures of atomic gases are listed
in~\cite{Hens}. Accuracy is better than the announced experimental
precision of 4\%.

\section{Structure formation}
\subsection{B\'{e}nard-Rayleigh thermal convection}
Stable vortices developing in a fluid bound by two horizontal plates at
different temperatures in a vertical (gravitational) force field  are
named after B'nard and Rayleigh. Since their first description in 1900
they have been the subject of an abundant literature, being a typical
example of structure formation in dynamic systems. The present discussion
aims at developing a set of differential equations based on Liouville's
fundamental equation~(\ref{liouville2}), where Hamiltonian iso-entropic
contributions and stochastic irreversible interactions are clearly
separated.

When convection develops in a fluid, the simple vertical symmetry defined
by the two temperature reservoirs and the external field acting
perpendicularly to the walls is broken. From unidimensional the problem
becomes bi- or tridimensional, with increased mathematical complexity.
Convective motion stimulated by thermal strain may show many different
patterns: rolls, cells, etc. The particular shape adopted by the system
depends primarily on lateral boundary conditions. Depending on symmetry,
we may want to have the coordinate system transformed. For simplicity we
consider generation of rolls, for which Cartesian coordinates are most
appropriate. Intensities referring to the different directions in space
will be indexed accordingly.

Let the fluid be confined between two parallel plates (distance $2D$,
taken as the $z-$direction) at different temperatures. The plates
represent a double temperature reservoir defining a given average inverse
temperature $\beta$ and a gradient $\gamma_z$.

The temperature gradient ($\gamma_z$) generates a linear moment of the
density distribution. This causes a non-vanishing value of $\theta_z$. The
relation between $\theta_z$ and $\gamma_z$ depends on conservation  of
momentum~(compare (\ref{mecha})). In the presence of gravity, pressure
gradient is balanced by the external force ($-mg$), whence~\cite{xh:92}:
\begin{equation}
\label{mecha_vert}
\theta_z - \frac{5}{2} \frac{\gamma_z}{\beta} = - \beta mgD.
\end{equation}

In field-free conditions, if $\gamma_z = 0$, the system's center of mass
is located half-way between the plates ($Z=0$). Thermal strain and the
external force displace the center of mass with respect to this neutral
position. Expressed as a function of the intensive variables and
using~(\ref{mecha_vert}), the vertical moment of the particle density
($N_z$) is given by~(\ref{Massieu}):
\begin{equation}
\label{N_Z}
N_z = \frac{D}{3}\;(\frac{\gamma_z}{\beta} - \beta mgD).
\end{equation}
Gravitation orients the gradient of the particle density to the bottom
regions but, if the system is heated from below, the distribution may
reverse. This situation presents much analogy with the population
inversion occurring in laser physics and the conclusions developed below
may readily be transferred to the domain of quantum
optics~\cite{xh:83,xh:85}.

With convection, intricate distributions develop, requiring for their
description additional or modified generating functions, supporting new
intensive variables. The distributions will be analyzed on the basis of
generating functions analogous to the set listed in
table~\ref{gen_conduction}, but extended and generalized to include the
additional constraints.

The geometric structure we want to focus on is periodic in the horizontal
direction. The macroscopic wavelength is $\lambda$, the rolls turning in
alternate directions. Let the $y-$axis connect successive vortices, the
$x-$direction being parallel to the motion's local global angular
momentum. We isolate along the $y-$direction a distance $\lambda/2$ in
which one vortex fits. This cell will be our system.

Let us focus on the intensities conjugate to the vertical and horizontal
moments of the particle density and investigate the role of the
source/sink contribution in Liouville's equation~(\ref{liouville2}). When
a vortex is active it perturbs the stationary distributions. The vertical
gradients are modified and horizontal gradients develop. Let the vertical
gradients dictated by the plates and valid in the absence of a vortex be
indexed $z,R$ from now on (e.g. $\theta_{z,R})$. Because of slow
relaxation of distortions due to finite transport in the fluid, in the
presence of a vortex this changes to $\theta_z$. Thermal diffusivity, a
process related to thermal conduction, tends to neutralize the latter
change. Relaxation of horizontal gradients (indexed $y$) follows the same
mechanism. The above may be summarized in the form of a set of linear
dissipation equations:
\begin{equation}
\begin{array}{l}
\dot{\theta_z} = - k_z (\theta_z - \theta_{z,R}) , \\
\dot{\theta_y} = - k_y \theta_y .
\label{dissipation}
\end{array}
 \end{equation}
with, considering the cell's geometry:
\begin{equation}
\begin{array}{l}
\label{diffusivity}
k_z = \left( \frac{\pi}{D} \right)^2 \kappa ,\\
k_y = \left[ \left( \frac{\pi}{2D} \right)^2 + \left(
\frac{2\pi}{\lambda} \right)^2 \right] \kappa ,
\end{array}
\end{equation}
($\kappa$: coefficient of thermal diffusivity).

In a second step we investigate the conservative iso-entropic part
of~(\ref{liouville2}).

Let $N_z$ and $N_y$ be the vertical and horizontal moments of the particle
density. Using~(\ref{N_Z}) and~(\ref{mecha_vert}) it is clear that $\d
N_z= \frac{2}{5} \d \theta_z$. The same argument holds in the horizontal
direction.

By referring to equation~(\ref{Gibbs}), considering possible changes of
the moments of the particle density, invariance of the entropy leads to:
\begin{equation}
\label{dS}
\d S = \beta mgD \d N_z - \theta_z \d N_z - \theta_y \d N_y = 0.
\end{equation}
where $mgD \d N_z$ represents change of the potential energy as the
system's centre of gravity moves vertically. By combining the two first
r.h.s. terms as $\theta_z^* = \theta_z - \beta mgD$, the constant entropy
condition may be rewritten $(\theta_z^*)^2 + \theta_y^2 =$~constant.

With $\omega$ representing the angular velocity characterizing the vortex,
iso-entropic circulation is described by the following set of differential
equations:
\begin{equation}
\begin{array}{l}
\dot{\theta_z^*} = - \omega \theta_y \, ,  \\
\dot{\theta_y} = \omega \theta_z^* \, .
\label{isoentropic}
\end{array}
\end{equation}

By combining the latter set with the equations for
dissipation~(\ref{dissipation}), the following set is obtained:
\begin{equation}
\begin{array}{l}
\label{bloch}
\dot{\theta_z^*} = - \omega \theta_y - k_z (\theta_z^* -
\theta_{z,R}^*)\,
, \\
\dot{\theta_y} = \omega \theta_z^* - k_y \theta_y.
\end{array}
\end{equation}
In laser physics, this set is named after Bloch.

Gravitation acting on the horizontal density gradient activates the vortex
while friction inhibits the collective motion. Complete balance between
the conflicting forces has been elaborated in~\cite{xh:92}. Full
development cannot be given here. It may however be drafted by:
\begin{equation}
\label{drive}
\dot{\omega} = \frac{2}{5} \frac{g}{D} \theta_y - \frac{G(\lambda)
\nu}{D^2} \omega \; ,
\end{equation}
where function $G(\lambda)$ depends on the form factor of the vortex.
$\nu$ is the viscosity of the fluid. By combining equations~(\ref{bloch})
and~(\ref{drive}), threshold inversion conditions for vortex formation and
the relevant form factor at and beyond threshold are easily determined
(pitchfork bifurcation). The results are in agreement with the
experiment~\cite{xh:92}. {\em Mutatis mutandis\/}, in laser physics, the
corresponding set of equations describes readily, next to the threshold
requirements, implications for bifurcations to a variety of unstable and
chaotic working conditions.

\subsubsection{Von Karman type turbulence}
In fluid dynamics it is known that the dimensionless number named after
Reynolds governs both turbulence in pipes or channels and the von Karman
vortex streets produced in flows past airfoils. This suggests that
analogous mechanisms may be at work in the latter different kinds of
systems. In order to stress the similarity, label {\em von Karman\/} might
be generalized by attaching it to turbulent flows in ducts as well.

The flows listed above have all been treated successfully
elsewhere~\cite{xh:92} in the same thermodynamic context. The detailed
mathematics being however rather cumbersome, only the principles involved
will be outlined next.

We consider a fluid flowing in a channel with constant width $2D$
(Poiseuille flow). In laminar conditions the velocity profile through the
channel is known to be quadratic. Contrasting with the Couette flow
discussed above, the table of generating functions required for the
thermodynamic description of the system implies therefore a quadratic
function for the collective momentum contribution. For the same reason,
the kinetic energy distribution involves a fourth order generating
function.

We are interested in the non-laminar regime. Instead of the above, let us
suppose therefore that a stream of vortices (with angular velocity
$\omega$) fitting exactly within the space between the walls flows down
the channel. The purpose is to examine this particular regime's stability.

Contrasting with the classical procedure where the walls are taken as
immobile, investigating the vortices implies the observer to be running
with the fluid at the same average speed ($v_y$). The walls are therefore
taken to be moving in the opposite direction.

The general dynamical equations for a vortex have been derived
above~(\ref{bloch}). In the present case there is however no external
force field to justify inversion. Symmetry breaking results however from
the asymmetric field of collective kinetic energy prevailing between the
channel's walls, caused by the very existence of the vortex superimposed
on local average downstream translational motion. Accurate analysis of the
relevant distribution of downstream collective kinetic energy indicates
that it contains at least a linear contribution (proportional to $z$, the
direction perpendicular to the walls). Its gradient is proportional to the
product $v_y \omega$.

Following Bernoulli's theorem, a gradient of collective kinetic energy in
a fluid generates a pressure or a density gradient perpendicularly to the
flow. This is equivalent to a force acting on the fluid, here in the $z-
$direction. This contains the product $v_y \omega$. It depends also on the
form factor of the relevant vortices. The modified set of dynamic
equations may therefore be written:
\begin{equation}
\begin{array}{l}
\label{bloch2}
\dot{\theta_z^*} = - \omega \theta_y - k_z [\theta_z^* -
C(\lambda)Dv_y \omega] , \\
\dot{\theta_y} = \omega \theta_z^* - k_y \theta_y.
\end{array}
\end{equation}
To the latter set an equation for $\dot{\omega}$ must be added.

The walls are the loci for mechanical interaction of the surroundings on
the system. Confining the fluid within its boundaries represents a force
exerted by the environment. Its magnitude is the local hydrodynamic
pressure. If the system is symmetric with respect to facing boundaries,
the forces exerted by the latter are equal and oppositely directed. In the
present case, asymmetric collective kinetic energy distribution with
respect to the boundaries (linear velocity superimposed on vortex) ensures
unequal coupling of the fluid with either walls.

When a particle collides with one of the walls, depending on whether this
is the one where the local average velocity gradient is higher or lower,
the reinjection trajectory following collision is more or less reoriented.
Hence, the force acting by the walls on the system is tangent to the flow
at one wall and perpendicular at the other wall. The perpendicular
pressure forces at either walls are not antagonistic. The resultant
(pressure = $n/\beta$) acts mechanically on the system. If the particle
density presents a gradient parallel to the walls ($\theta_y \neq 0$) the
resulting moment of the forces activates rotation of the vortex. Hence:
\begin{equation}
\label{drive2}
\dot{\omega} = \frac{1}{4 \beta mD^2} \theta_y - \frac{G(\lambda)
\nu}{D^2} \omega\;.
\end{equation}

In stationary conditions, the dotted functions vanish. By eliminating
$\theta_z$ and $\theta_y$ in the resulting set of equations, the following
equation for $\omega$ results:
\begin{equation}
\label{vortex}
\omega^2 - \frac{k_zC}{8G\beta m }\;{\rm Re}\;\omega + k_y k_z = 0 \; .
\end{equation}
where ${\rm Re} = 2Dv_y/\nu$ is Reynold's number.

Besides the trivial solution ($\omega = 0$), expression~(\ref{vortex}) is
the characteristic equation for vortex stability. It has two roots (limit
point bifurcation). Depending on the value of the Reynolds number they may
be real or complex. Only real solutions justify stable vortices. The value
of the critical Reynolds number separating conditions for stable and
unstable vortices depends on their form factor. The lowest value must be
retained. It has been calculated in~\cite{xh:92} for flow in a channel, in
a square section pipe and past an airfoil. Results are in agreement with
the experiment.

\section{Conclusions}
The present paper reviews the state of affairs for solving the
irreversibility paradox. The conflict raises from apparent
contradiction between reversibility of microscopic laws of motion and the
irreversible behaviour of macroscopic systems. The analysis convincingly
demonstrates that attempts based on the assumption that the relevant
macroscopic systems are perfectly isolated, cannot justify the law of
increase of entropy. Such systems are indeed necessarily conservative, a
property that holds for the entropy too. Some of the attempts towards
escaping this fundamental symmetry property have been discussed. Their
arguments have been shown to be inconclusive. As a result one is forced us
to assume that perfect isolation conditions are incompatible with real
physical systems.

The notion of quasi-isolation has been introduced, indicating the
condition of systems that are allowed to exchange energy fluctuations with
their environment. With closed systems, the environment is at least for a
part represented by their walls.

The conceptual basis for the theoretical investigation of the dynamics of
quasi-isolated systems is furnished by the experimental evidences obtained
from the analysis of Joule's experiment.

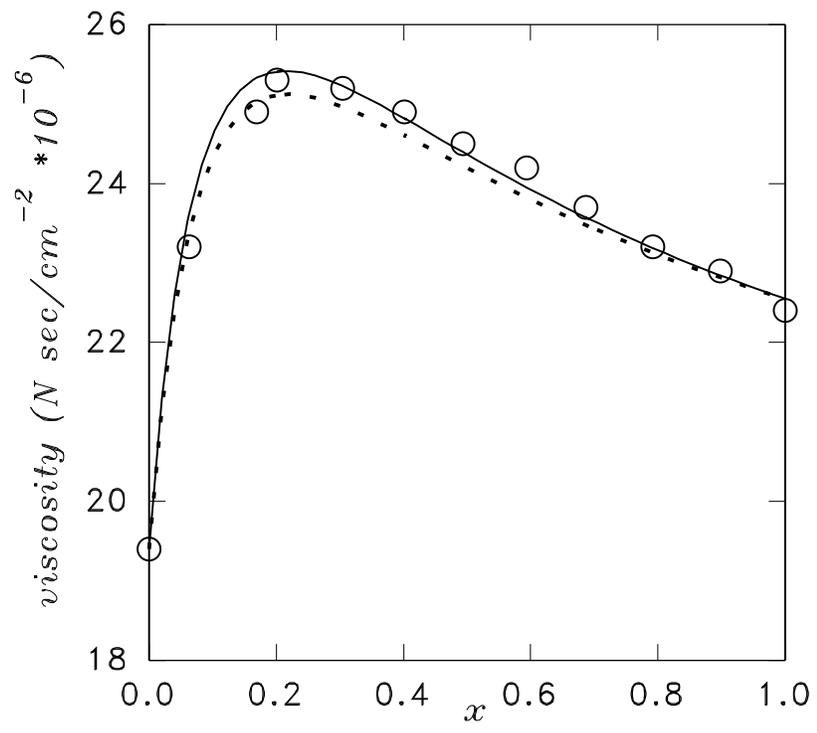
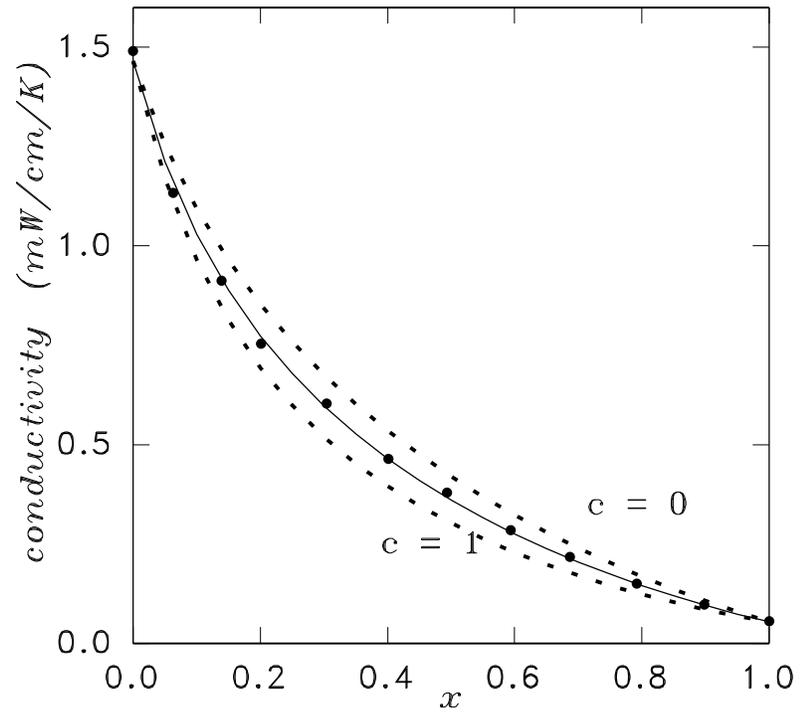
\end{document}